\documentstyle[12pt]{article}
\def\la{\mathrel{\mathpalette\fun <}}
\def\ga{\mathrel{\mathpalette\fun >}}
\def\fun#1#2{\lower3.6pt\vbox{\baselineskip0pt\lineskip.9pt
\ialign{$\mathsurround=0pt#1\hfil##\hfil$\crcr#2\crcr\sim\crcr}}}

\begin{document}
\topmargin -0.5cm
\oddsidemargin -0.3cm

\begin{titlepage}
\pagestyle{empty}
\begin{flushright}
{ITEP-TH-55/96}
\end{flushright}
\vspace*{1mm}
\begin{center}
{\bf TESTS OF CPT}\footnote{Talk at the Workshop on $K$-physics,
ORSAY, France, 30 May - 4 June, 1996.}

\vspace{0.3in}
{\bf L.B.  Okun}\\
\vspace{0.1in}
ITEP, Moscow, 117218, Russia\\
\vspace{0.5in}
 {\bf Abstract \\}
 \end{center}
A few remarks concerning theoretical suggestions and experimental
tests of CPT during 1980's - 1990's. Is it worth to search for the
particle-antiparticle mass differences in sectors other than
$K^0\bar{K}^0$?

\end{titlepage}

\vfill\eject
\pagestyle{empty}

\setcounter{page}{1}
\pagestyle{plain}

\section{Progress in ${\bf K^0, \bar{K}^0}$.}

The main activity on testing CPT has been concentrated on the
experimental study and theoretical analysis of neutral kaons. In the
early 1980's, following the lines of ref. \cite{1}, it was realized
\cite{2,3} that the large difference between $\phi_{00}$ on the
one side and $\phi_{+-}$ and $\phi_{SW}$ on the other indicated
a $2\sigma$ discrepancy with CPT. Special experiments at CERN (NA31)
\cite{4} and FNAL (E731) \cite{5} have wiped out the discrepancy. The
central values of $\phi_{00}$ and $\phi_{+-}$ coincided, while the
uncertainties have been reduced to $2^0$. At present (see the talks
by R.Briere (FNAL, E773), R. Le Gac (CPLEAR), and P.Pavlopoulos
(CPLEAR) at this Workshop) $\phi_{+-}$,  $\phi_{00}$ and $\phi_{SW}$
agree with accuracy of the order of $1^0$. The CPLEAR with its tagged
$K^0$'s and $\bar{K}^0$'s was of special importance in this respect.

Both the CPLEAR, which already finishes its life, and the DA$\Phi$NE,
which only starts its life, have played an important role by inspiring
theorists to perform more detailed phenomenological analysis of the
CPT tests in the $K^0 \bar{K}^0$ system. In particular, to take into
account the possible violation of CPT in the semileptonic decays, not
to rely on the Bell-Steinberger unitarity relation, to stress the
necessity to test $T$-invariance of CP-even terms \cite{6,7,8}.

Another source of inspiration has been provided by supergravity and
superstrings ("vacuum foam" and loss of unitarity, breaking of
Lorentz-invariance and hence of CPT, see talks by J.Ellis, P.Huet
and A.Kostelecky at this Workshop).

\section{CPT tests outside ${\bf K^0, \bar{K}^0}$.}

If we are lucky and there exists interaction which violates baryon
number conservation by two units, then there may arise a test of
CPT, which is even more sensitive then the mass difference of $K^0$
and $\bar{K}^0$. To test CPT one has to compare two phenomena: 1) the
decay of $O^{16}$ into hadrons with total $A=14$ and with energy
release $\sim 1.9$ GeV; 2) the vacuum transition of a neutron into
antineutron. Both phenomena have been considered in a number of
theoretical papers \cite{10} - \cite{21}. The lower limit for the
period of $n - \bar{n}$ oscillations has been established at
ILL-Grenoble \cite{22}:
$$
\tau_{n\bar{n}} > 8.6 \cdot 10^7 \mbox{\rm sec.}
$$
The lower limit for lifetime of $O^{16}$, according to Kamiokande
\cite{23},
$$
\tau(O^{16}) > 2.4 \cdot 10^{31} \mbox{\rm yr.}
$$
(Note that from Frejus experiment \cite{24}:
$$
\tau(Fe) > 6.5 \cdot 10^{31} \mbox{\rm yr.}
$$

We discuss here $O^{16}$ because in a not too distant future its
stability will be tested with much higher precision at Super
Kamiokande \cite{25}.)

There is an Oak Ridge proposal \cite{26, 27} to increase the lower
limit for $\tau_{n\bar{n}}$ up to $10^{10}$ sec.

There is no unanimity among theorists in extracting the limit on
$n - \bar{n}$ oscillations from the existing data of Kamiokande
\cite{23} and Frejus \cite{24}. For example, $\tau_{n\bar{n}}\ga
10^8$ sec is predicted in ref. \cite{17} and \cite{19}, while
$\tau_{n\bar{n}}\ga 10^9$ sec is predicted in ref. \cite{21}. Let us
optimistically assume that a consensus will be reached and a reliable
theoretical relation between $\tau(O^{16})$ and $\tau_{n\bar{n}}$
will be established. Moreover let us imagine that the decay of
$O^{16}$ with 1.9 GeV release is discovered by the Super Kamiokande
physicists. Then the observation of $n-\bar{n}$ oscillations would
put a stringent limit on the $n-\bar{n}$ mass difference, while the
non-observation may mean that $m_n \neq m_{\bar{n}}$.

The search of $n-\bar{n}$ transitions in vacuum is carried out in an
intense beam of ultracold neutrons. The beam is moving in vacuum,
being screened from magnetic field of the earth, which otherwise
would remove the energy degeneracy between $n$ and $\bar{n}$,
because of opposite signs of their magnetic moments. After a
free-flight time $t$, of the order of $0.1-0.01$ sec, the beam hits a
target. If antineutrons appear during this time, they should
annihilate in the target releasing $\sim 1.9$ GeV of energy in the
form of mesons, photons, recoil nucleons etc. The probability that
$n$ will transform into $\bar{n}$ during time $t$ is equal to
$(t/\tau_{n\bar{n}})^2$. Therefore intense neutron beams are needed.

Now let us return to CPT. If $m_n \neq m_{\bar{n}}$, then the
transitions $n\to \bar{n}$ could be observed, only if $\Delta
m_{n\tilde{n}}\la 1/t$. Thus, as noticed in ref. \cite{28}, if $n\to
\bar{n}$ transitions are observed, that puts an upper limit on the
mass difference $\Delta m_{n\bar{n}}/m_n \la 10^{-22} - 10^{-23}$
which is by 4-5 orders of magnitude better than the existing limit on
$\Delta m_{K\bar{K}}$.

If $\Delta m_{n\bar{n}} \gg 1/t$, then $n-\bar{n}$ transitions are
strongly suppressed by a small factor ($t\cdot \Delta
m_{n\bar{n}})^{-2}$ and become unobservable. The proof of CPT
violation in this case would be the observation of decays of $O^{16}$
at Super Kamiokande discussed above. The tiny mass difference $\Delta
m_{n\bar{n}}$ has no influence on the rate of $O^{16}$ decays. Note
that $\Delta m_{n\bar{n}}/m_n \sim \frac{m_n}{M_{Planck}} \sim
10^{-19}$ is expected in superstring and supergravity inspired
speculations.

The question of other particle-antiparticle mass differences is often
addressed: $m_{\mu^+} - m_{\mu^-}$, $m_{e^+} - m_{e^-}$,
$m_{\pi^+} - m_{\pi^-}$, $m_{K^+} - m_{K^-}$, $m_p - m_{\bar{p}}$.
The accuracy of their measurements will never reach the accuracy of
$m_{K^0} - m_{\bar{K}^0}$. Is it worth to continue to measure them?
The answer is yes! There may exist certain selection rules for the
CPT-violating interaction, so that $\Delta m_K$ may be not sensitive
to $\Delta m_{\mu}$ and $\Delta m_e$. Even for the charged pions and
kaons, as well as protons, one may say that they, unlike neutral
kaons, contain $u$-quarks. Thus, the mass difference between $u$ and
$\bar{u}$ may not so strongly manifest itself in neutral kaons.
This argument is not absolutely convincing because of existence of
weak interactions, which mixes various flavours of quarks.

But, irrespective of all these theoretical considerations, one has
to follow the advice of Galileo and measure everything that can be
measured.

\end{document}